\newif\ifpagetitre            \pagetitretrue
\newtoks\hautpagetitre        \hautpagetitre={\hfil}
\newtoks\baspagetitre         \baspagetitre={\hfil}
\newtoks\auteurcourant        \auteurcourant={\hfil}
\newtoks\titrecourant         \titrecourant={\hfil}

\newtoks\hautpagegauche       \newtoks\hautpagedroite
\hautpagegauche={\hfil\the\auteurcourant\hfil}
\hautpagedroite={\hfil\the\titrecourant\hfil}

\newtoks\baspagegauche \baspagegauche={\hfil\tenrm\folio\hfil}
\newtoks\baspagedroite \baspagedroite={\hfil\tenrm\folio\hfil}

\headline={\ifpagetitre\the\hautpagetitre
\else\ifodd\pageno\the\hautpagedroite
\else\the\hautpagegauche\fi\fi}

\footline={\ifpagetitre\the\baspagetitre
\global\pagetitrefalse
\else\ifodd\pageno\the\baspagedroite
\else\the\baspagegauche\fi\fi}

\vsize=9.0in\voffset=1cm
\looseness=2


\message{fonts,}

\font\tenrm=cmr10
\font\ninerm=cmr9
\font\eightrm=cmr8
\font\teni=cmmi10
\font\ninei=cmmi9
\font\eighti=cmmi8
\font\ninesy=cmsy9
\font\tensy=cmsy10
\font\eightsy=cmsy8
\font\tenbf=cmbx10
\font\ninebf=cmbx9
\font\tentt=cmtt10
\font\ninett=cmtt9

\font\ninesl=cmsl9
\font\eightsl=cmsl8

\font\nineit=cmti9
\font\eightit=cmti8

\skewchar\ninei='177 \skewchar\eighti='177
\skewchar\ninesy='60 \skewchar\eightsy='60

\def\eightpoint{\def\rm{\fam0\eightrm} 
\normalbaselineskip=9pt
\normallineskiplimit=-1pt
\normallineskip=0pt

\textfont0=\eightrm \scriptfont0=\sevenrm \scriptscriptfont0=\fiverm
\textfont1=\ninei \scriptfont1=\seveni \scriptscriptfont1=\fivei
\textfont2=\ninesy \scriptfont2=\sevensy \scriptscriptfont2=\fivesy
\textfont3=\tenex \scriptfont3=\tenex \scriptscriptfont3=\tenex
\textfont\itfam=\eightit  \def\it{\fam\itfam\eightit} 
\textfont\slfam=\eightsl \def\sl{\fam\slfam\eightsl} 

\setbox\strutbox=\hbox{\vrule height6pt depth2pt width0pt}%
\normalbaselines \rm}

\def\ninepoint{\def\rm{\fam0\ninerm} 
\textfont0=\ninerm \scriptfont0=\sevenrm \scriptscriptfont0=\fiverm
\textfont1=\ninei \scriptfont1=\seveni \scriptscriptfont1=\fivei
\textfont2=\ninesy \scriptfont2=\sevensy \scriptscriptfont2=\fivesy
\textfont3=\tenex \scriptfont3=\tenex \scriptscriptfont3=\tenex
\textfont\itfam=\nineit  \def\it{\fam\itfam\nineit} 
\textfont\slfam=\ninesl \def\sl{\fam\slfam\ninesl} 
\textfont\bffam=\ninebf \scriptfont\bffam=\sevenbf
\scriptscriptfont\bffam=\fivebf \def\bf{\fam\bffam\ninebf} 
\textfont\ttfam=\ninett \def\tt{\fam\ttfam\ninett} 

\normalbaselineskip=11pt
\setbox\strutbox=\hbox{\vrule height8pt depth3pt width0pt}%
\let \smc=\sevenrm \let\big=\ninebig \normalbaselines
\parindent=1em
\rm}

\def\tenpoint{\def\rm{\fam0\tenrm} 
\textfont0=\tenrm \scriptfont0=\ninerm \scriptscriptfont0=\fiverm
\textfont1=\teni \scriptfont1=\seveni \scriptscriptfont1=\fivei
\textfont2=\tensy \scriptfont2=\sevensy \scriptscriptfont2=\fivesy
\textfont3=\tenex \scriptfont3=\tenex \scriptscriptfont3=\tenex
\textfont\itfam=\nineit  \def\it{\fam\itfam\nineit} 
\textfont\slfam=\ninesl \def\sl{\fam\slfam\ninesl} 
\textfont\bffam=\ninebf \scriptfont\bffam=\sevenbf
\scriptscriptfont\bffam=\fivebf \def\bf{\fam\bffam\tenbf} 
\textfont\ttfam=\tentt \def\tt{\fam\ttfam\tentt} 

\normalbaselineskip=11pt
\setbox\strutbox=\hbox{\vrule height8pt depth3pt width0pt}%
\let \smc=\sevenrm \let\big=\ninebig \normalbaselines
\parindent=1em
\rm}

\message{fin format jgr}

\hautpagegauche={\hfill\ninerm\the\auteurcourant}
\hautpagedroite={\ninerm\the\titrecourant\hfill}
\auteurcourant={R.G.\ Novikov}
\titrecourant={Formulas  for phase recovering from phaseless
scattering data at fixed frequency}

\magnification=1200
\font\Bbb=msbm10
\def\R{\hbox{\Bbb R}}

\def\v{\varphi}

\vskip 2 mm
\centerline{\bf Phaseless  inverse scattering in the one-dimensional case}

\vskip 2 mm
\centerline{\bf R.G.\ Novikov}
\vskip 2 mm

\noindent
{\ninerm CNRS (UMR 7641), Centre de Math\'ematiques Appliqu\'ees, Ecole
Polytechnique,}

\noindent
{\ninerm 91128 Palaiseau, France;}

\noindent
{\ninerm IEPT RAS, 117997 Moscow, Russia}


\noindent
{\ninerm e-mail: novikov@cmap.polytechnique.fr}

\vskip 2 mm
{\bf Abstract.}
We consider  the one-dimensional Schr\"odinger equation with a potential satisfying the standard assumptions
of the inverse scattering theory and supported on the half-line $x\ge 0$. For this equation at fixed positive
energy we give explicit formulas for finding the full complex valued reflection coefficient to the left
from appropriate  phaseless  scattering data measured on the left, i.e. for $x<0$.
Using these formulas and known inverse scattering results we obtain global uniqueness and
reconstruction results for  phaseless  inverse scattering in dimension $d=1$.

\vskip 2 mm
{\bf 1. Introduction}

We consider the one-dimensional Schr\"odinger  equation
$$-{d^2\over dx^2}\psi +v(x)\psi=E\psi,\ \ x\in\R,\ \ E>0,\eqno(1.1)$$
where  $v$  satisfies the standard  assumptions of the  inverse scattering theory (see [F]) and is
supported on the half-line $x\ge 0$. More precisely, we assume that
$$\eqalign{
&v\ \ {\rm is\ real-valued},\ \ v\in  L^1_1(\R),\cr
&v(x)\equiv 0\ \ {\rm for}\ \ x<0,\cr}\eqno(1.2)$$
where
$$L_1^1(\R)=\{u\in L^1(\R):\ \int\limits_{\R}(1+|x|)|u(x)|dx<\infty\}.\eqno(1.3)$$
For equation (1.1) we consider the scattering solution $\psi^+=\psi^+(\cdot,k)$, $k=\sqrt{E}>0$,
continuous and bounded on $\R$ and specified by the following asymptotics:
$$\psi^+(x,k)=\left\{\matrix{
e^{ikx}+s_{21}(k)e^{-ikx}\ \ &\ {\rm as}\ \ x\to -\infty,\hfill\cr
s_{22}(k)e^{ikx}+o(1)\ \ &\ {\rm as}\ \ x\to +\infty,\hfill\cr}\right.\eqno(1.4)$$
for some a priori unknown $s_{21}$ and $s_{22}$.  In addition, the coefficients $s_{21}$ and $s_{22}$
arising in (1.4) are the reflection coefficient to the left and transmission  coefficient to the right,
respectively, for equation (1.1).

In order to find $\psi^+$ and $s_{21}$,  $s_{22}$ from $v$ one can use well-known results of the
one-dimensional direct  scattering theory, see e.g. [F]. And properties of $\psi^+$, $s_{21}$,  $s_{22}$
are known in detail, see [DT], [F], [HN], [L].

In particular, it is well-known that
$$|s_{21}(k)|^2+|s_{22}(k)|^2=1,\ \ |s_{21}(k)|^2<1,\ \ k>0.\eqno(1.5)$$

Let
$$\R_+=]0,+\infty [,\ \ \R_-=] -\infty,0 [.\eqno(1.6)$$

We consider the following two types of scattering data measured on the left for equation  (1.1): (a) $s_{21}(k)$ and
(b) $\psi^+(x,k)$, \ \ $x\in X_-\subseteq\R_-$,
where $k=\sqrt{E}>0$.

In addition, we consider the following inverse scattering problems:

{\bf Problem 1.1a.}
 Reconstruct potential $v$ on $\R$ from its reflection coefficient $s_{21}$ on $\R_+$.

{\bf Problem 1.1b.}
 Reconstruct potential $v$ on $\R$ from its scattering data
 $\psi^+(x,\cdot)$ on $\R_+$ at fixed $x\in\R_-$.

{\bf Problem 1.2a.}
 Reconstruct potential $v$ on $\R$ from  its phaseless scattering data $|s_{21}|^2$ on $\R_+$.

{\bf Problem 1.2b.}
 Reconstruct potential $v$ on $\R$ from its phaseless scattering data
 $|\psi^+|^2$ on $X_-\times\R_+$ for some appropriate $X_-$.

{\bf Problem 1.2c.}
 Reconstruct potential $v$ on $\R$ from its phaseless scattering data
$|s_{21}|^2$ on $\R_+$ and $|\psi^+|^2$ on $X_-\times\R_+$ for some appropriate $X_-$.

Note that in quantum mechanical scattering experiments in the framework of model described by
equation (1.1) the   phaseless scattering data $|s_{21}|^2$,  $|\psi^+|^2$ of Problems 1.2a-1.2c
can be measured directly, whereas the complex-valued scattering data $s_{21}$, $\psi^+$
of Problems 1.1a, 1.1b are not accessible for direct measurements.
Therefore, Problems 1.2 are of particular applied interest in the framework of inverse scattering
of quantum mechanics. However, Problems 1.1 are much more considered in the literature than Problems 1.2.
See  [ChS], [DT], [F], [L], [M], [N1], [NM] and references therein in connection with Problem 1.1a and
[AS], [KS] in connection with Problem 1.2a and its modification.

In particular, work [NM] gives global uniqueness and reconstruction results for Problem 1.1a;
see also [AW], [GS] and references given in [AW].
 And , obviously, Problem 1.1b is reduced to Problem 1.1a by the formula
$$s_{21}(k)=e^{ikx}\psi^+(x,k)-e^{2ikx},\ \ x\in\R_-,\ \ k\in\R_+.\eqno(1.7)$$

On the other hand, for Problem 1.2a it is well known that the phaseless scattering data
$|s_{21}|^2$ on $\R_+$ do not determine $v$ uniquely, in general. In particular,  we have that
$$\eqalign{
&s_{21,y}(k)=e^{2iky}s_{21}(k),\cr
&|s_{21,y}(k)|^2=|s_{21}(k)|^2,\ \ k\in\R_+,\ \ y\in\R,\cr}\eqno(1.8)$$
where $s_{21}$ is the reflection coefficient to the left for $v$ and $s_{21,y}$ is
the reflection coefficient to the left for $v_y$, where
$$v_y(x)=v(x-y),\ \ x\in\R,\ \ y\in\R.\eqno(1.9)$$

In the present work we continue studies of [N3]. We recall that article [N3] gives, in particular,
explicit asymptotic formulas for finding the complex-valued scattering amplitude from appropriate
phaseless scattering data for the Schr\"odinger equation  at fixed energy in dimension $d\ge 2$.
Then using these formulas related phaseless inverse scattering problems are reduced in [N3] in
dimension $d\ge 2$ to well developed inverse scattering from the complex-valued scattering amplitude.
And, as a corollary, [N3] contains different results on inverse scattering without phase information
in dimension $d\ge 2$.

In connection with recent results on phaseless inverse scattering in dimension $d\ge 2$, see also
[K], [KR1], [KR2], [N2] and references therein.

In order to present results of the present work we use the notations:
$$\eqalign{
&S_1(x_1,x_2,k)=\{|s_{21}(k)|^2,|\psi^+(\cdot,k)|^2\ \ {\rm on}\ \ X_-\},\cr
&{\rm where}\ \ X_-=\{x_1,x_2\in\R_-:\ \ x_1\ne x_2\},\ \ k\in\R_+;\cr}\eqno(1.10)$$
$$\eqalign{
&S_2(x_1,x_2,x_3,k)=|\psi^+(\cdot,k)|^2\ \ {\rm on}\ \ X_-,\cr
&{\rm where}\ \ X_-=\{x_1,x_2,x_3\in\R_-:\ \ x_i\ne x_j\ \ {\rm if}\ \ i\ne j\},\ \ k\in\R_+;\cr}\eqno(1.11)$$
$$S_3(x,k)=\{|\psi^+(x,k)|^2,{d|\psi^+(x,k)|^2\over dx}\}, x\in\R_-,\  k\in\R_+.\eqno(1.12)$$

Using these notations the main results of the present work can be summarized as follows:

(A1) We give explicit formulas for finding $s_{21}(k)$ from $S_1(x_1,x_2,k)$ for fixed $x_1$, $x_2$
and $k$, where $x_1\ne x_2\ mod(\pi(2k)^{-1})$; see Theorem 2.1 of Section 2.

(A2) We give explicit formulas for finding $s_{21}(k)$ from $S_2(x_1,x_2,x_3,k)$ for fixed $x_1$, $x_2$, $x_3$
and $k$, where $x_i\ne x_j\ mod(\pi k^{-1})$ if $i\ne j$; see Theorem 2.2 of Section 2.

(A3) We give explicit formulas for finding $s_{21}(k)$ from $S_3(x,k)$ for fixed $x$
and $k$; see Theorem 2.3 of Section 2.

(B)  We give global uniqueness and reconstruction results (1) for finding $v$ on $\R$ from $S_1(x_1,x_2,\cdot)$
on $\R_+$  for fixed $x_1$, $x_2$, (2)  for finding $v$ on $\R$ from $S_2(x_1,x_2,x_3,\cdot)$
on $\R_+$  for fixed $x_1$, $x_2$, $x_3$, and (3)  for finding $v$ on $\R$ from $S_3(x,\cdot)$
on $\R_+$ at fixed $x$; see Theorem 2.4 of Section 2.

Note that results of (B1)-(B3) follow from (A1)-(A3) and the aforementioned results of [NM] on Problem 1.1a. In addition,
the results of (B1) are global results on Problem 1.2c and the results of (B2), (B3)  are global results on Problem 1.2b.

The main results of the present work are presented in detail in Section 2.

\vskip 2 mm
{\bf 2. Main results}

We represent $s_{21}$ of (1.4) as follows:
$$s_{21}(k)=|s_{21}(k)|e^{i\alpha(k)},\ \ k\in\R_+.\eqno(2.1)$$
We consider
$$
a(x,k)=|\psi^+(x,k)|^2-1,\ \ x\in\R_-,\ \ k\in\R_+,\eqno(2.2)$$
where $\psi^+$ is the scattering solutions of (1.4).

\vskip 2 mm
{\bf Theorem 2.1.}
{\sl
Let potential $v$ satisfy (1.2) and $s_{21}$, $a$ be the functions of (1.4), (2.2). Let
$x_1,x_2\in\R_-$, $k\in\R_+$, $x_1\ne x_2\ mod(\pi(2k)^{-1})$. Then:
$$\eqalign{
&|s_{21}|\pmatrix{
\cos\alpha\cr
\sin\alpha\cr}=(2\sin(2k(x_2-x_1)))^{-1}\times\cr
&\pmatrix{
\sin(2kx_2)\ &\ -\sin(2kx_1)\cr
-\cos(2kx_2)\ &\ \cos(2kx_1)\cr}
\pmatrix{
a(x_1,k)-|s_{21}|^2\cr
a(x_2,k)-|s_{21}|^2\cr},
\cr}\eqno(2.3)$$
where $|s_{21}|=|s_{21}(k)|$, $\alpha=\alpha(k)$.
}

Theorem 2.1 is proved in Section 3.

One can see that Theorem 2.1 gives explicit formulas for finding $s_{21}(k)$ from

\noindent
$S_1(x_1,x_2,k)$ of (1.10) for fixed $x_1$, $x_2$ and $k$, where $x_1\ne x_2\ mod(\pi(2k)^{-1})$.

\vskip 2 mm
{\bf Theorem 2.2.}
{\sl
Let potential $v$ satisfy (1.2) and $s_{21}$, $a$ be the functions of (1.4), (2.2). Let
$x_1,x_2,x_3\in\R_-$, $k\in\R_+$, $x_1\ne x_2\ mod(\pi k^{-1})$ if $i\ne j$. Then:
$$\eqalign{
&|s_{21}|\pmatrix{
\cos\alpha\cr
\sin\alpha\cr}=(8(\sin(k(x_2-x_3))\sin(k(x_2-x_1))\sin(k(x_1-x_3)))^{-1}\times\cr
&\pmatrix{
\sin(2kx_3)-\sin(2kx_1)\ &\ -\sin(2kx_2)+\sin(2kx_1)\cr
-\cos(2kx_3)+\cos(2kx_1)\ &\ \cos(2kx_2)-\cos(2kx_1)\cr}
\pmatrix{
a(x_2,k)-a(x_1,k)\cr
a(x_3,k)-a(x_1,k)\cr},\cr}\eqno(2.4)$$
where $|s_{21}|=|s_{21}(k)|$, $\alpha=\alpha(k)$.
}

Theorem 2.2 is proved in Section 3.

One can see that Theorem 2.2 gives explicit formulas for finding $s_{21}(k)$ from

\noindent
$S_2(x_1,x_2,x_3,k)$ of (1.11) for fixed $x_1$, $x_2$, $x_3$ and $k$, where $x_1\ne x_2\ mod(\pi k^{-1})$ if $i\ne j$.

\vskip 2 mm
{\bf Theorem 2.3.}
{\sl
Let potential $v$ satisfy (1.2) and $s_{21}$, $\psi^+$ be the functions of (1.4).
Then the following formulas hold:
$$\eqalignno{
&Re\,(s_{21}(k)e^{-ikx})=-1+(|\psi^+(x,k)|^2-|Im\,(s_{21}(k)e^{-ikx})|^2)^{1/2},&(2.5)\cr
&Im\,(s_{21}(k)e^{-ikx})={1\over 4k}{d|\psi^+(x,k)|^2\over dx},&(2.6)\cr}$$
where $x\in\R_-$, $k\in\R_+$, and  $(\cdot)^{1/2}>0$ in (2.5).
}

Theorem 2.3 is proved in Section 3.

One can see that Theorem 2.3 gives explicit formulas for finding $s_{21}(k)$ from
$S_3(x,k)$ of (1.12).

As corollaries of Theorems 2.1, 2.2, 2.3 and results of [NM], we obtain
the following global uniqueness and reconstruction results on phaseless inverse scattering
for equation (1.1):

{\bf Theorem 2.4.}
{\sl
Let potential $v$ satisfy (1.2) and $S_1$, $S_2$, $S_3$ be the phaseless scattering data
of (1.10), (1.11), (1.12). Then: (1) $S_1(x_1,x_2,\cdot)$ on $\R_+$, for fixed $x_1$, $x_2$,
uniquely determine $v$ on $\R$ via formulas (2.1)-(2.3) and results of [NM];
(2) $S_2(x_1,x_2,x_3,\cdot)$ on $\R_+$, for fixed $x_1$, $x_2$, $x_3$,
uniquely determine $v$ on $\R$ via formulas (2.1), (2.2), (2.4) and results of [NM];
(3) $S_3(x,\cdot)$ on $\R_+$, for fixed $x$,
uniquely determine $v$ on $\R$ via formulas (2.5), (2.6) and results of [NM].

}

\vskip 2 mm
{\bf 3. Proofs of Theorems 2.1, 2.2 and 2.3}
\vskip 2 mm
{\it Proof of Theorem 2.1.}
Using (1.4) we obtain that
$$\eqalign{
&|\psi^+(x,k)|^2=\psi^+(x,k)\overline{\psi^+(x,k)}=\cr
&1+2Re\,(s_{21}(k)e^{-2ikx})+|s_{21}(k)|^2,\ \ x\in\R_-,\ \ k\in\R_+.\cr}\eqno(3.1)$$
In addition, in view of (2.1) we have that
$$2Re\,(s_{21}(k)e^{-2ikx})=2|s_{21}(k)|\cos(2kx-\alpha(k)).\eqno(3.2)$$
Using (2.2), (3.1), (3.2) we obtain that
$$\eqalign{
&|s_{21}(k)|\cos(2kx)\cos(\alpha(k))+\sin(2kx)\sin(\alpha(k))=\cr
&2^{-1}(a(x,k)-|s_{21}(k)|^2),\ \  x\in\R_-,\ \ k\in\R_+.\cr}\eqno(3.3)$$
Using (3.3) for $x=x_1$ and $x=x_2$, we obtain the system
$$\eqalign{
&\pmatrix{
\cos(2kx_1)\ &\ \sin(2kx_1)\cr
\cos(2kx_2)\ &\ \sin(2kx_2)\cr}
|s_{21}|\pmatrix{
\cos\alpha\cr
\sin\alpha\cr}=\cr
&2^{-1}\pmatrix{
a(x_1,k)-|s_{21}|^2\cr
a(x_2,k)-|s_{21}|^2\cr},
\cr}\eqno(3.4)$$
where $|s_{21}|=|s_{21}(k)|$, $\alpha=\alpha(k)$.

Formula (2.3) follows from (3.4).

Theorem 2.1 is proved.

\vskip 2 mm
{\it Proof of Theorem 2.2.}
Subtracting equality (3.3) for $x=x_1$ from equality (3.3) for $x=x_2$ and
from equality (3.3) for $x=x_3$, we obtain the system
$$\eqalign{
&\pmatrix{
\cos(2kx_2)-\cos(2kx_1)\ &\ \sin(2kx_2)-\sin(2kx_1)\cr
\cos(2kx_3)-\cos(2kx_1)\ &\ \sin(2kx_3)-\sin(2kx_1)\cr}
|s_{21}|\pmatrix{
\cos\alpha\cr
\sin\alpha\cr}=\cr
&2^{-1}\pmatrix{
a(x_2,k)-a(x_1,k)\cr
a(x_3,k)-a(x_1,k)\cr},
\cr}\eqno(3.5)$$
where $|s_{21}|=|s_{21}(k)|$, $\alpha=\alpha(k)$.

One can see that
$$\Delta=\sin(2k(x_3-x_2))+\sin(2k(x_2-x_1))+\sin(2k(x_1-x_3)),\eqno(3.6)$$
where $\Delta$ is the determinant of the system (3.5). In addition, using the formulas
$$\eqalign{
&\sin\v_1+\sin\v_2=2\cos\bigl({{\v_1-\v_2}\over 2}\bigr)\sin\bigl({{\v_1+\v_2}\over 2}\bigr),\cr
&\sin(\v_1+\v_2)=2\cos\bigl({{\v_1+\v_2}\over 2}\bigr)\sin\bigl({{\v_1+\v_2}\over 2}\bigr),\cr
&\sin\v_1+\sin\v_2-\sin(\v_1+\v_2)=4\sin\bigl({{\v_1+\v_2}\over 2}\bigr)\sin\bigl({\v_1\over 2}\bigr)\sin\bigl({\v_2\over 2}\bigr)\cr}
\eqno(3.7)$$
for $\v_1=2k(x_2-x_1)$, $\v_2=2k(x_1-x_3)$,
we obtain
$$\Delta=4\sin(k(x_2-x_3))\sin(k(x_2-x_1))\sin(k(x_1-x_3)).\eqno(3.8)$$

Formula (2.4) follows from (3.5), (3.8).

Theorem 2.2 is proved.

\vskip 2 mm
{\it Proof of Theorem 2.3.}
Using (3.1) we obtain that
$$(Re\,(s_{21}(k)e^{-2ikx})+1)^2+(Im\,(s_{21}(k)e^{-2ikx}))^2=|\psi^+(x,k)|^2.\eqno(3.9)$$
Formula (2.5), where $(\cdot)^{1/2}>0$, follows from (3.9) and the property that $|s_{21}(k)|<1$,
see (1.5).

Using (2.1), (3.1), (3.2) we obtain that
$${d|\psi^+(x,k)|^2\over dx}=4k|s_{21}(k)|\sin\,(\alpha(k)-2kx)=4kIm\,(s_{21}(k)e^{-2ikx}).\eqno(3.10)$$
Formula (2.6) follows from (3.10).

Theorem 2.3 is proved.


\vskip 4 mm
{\bf References}
\vskip 2 mm
\item{[ AS]} T. Aktosun, P.E. Sacks, Inverse problem on the line without phase information,
Inverse Problems 14, 1998, 211-224.
\item{[ AW]} T. Aktosun, R. Weder, Inverse scattering with partial information on the potential,
J. Math. Anal. Appl. 270, 2002, 247-266.
\item{[ChS]} K. Chadan, P.C. Sabatier, Inverse Problems in Quantum Scattering
Theory, 2nd edn. Springer, Berlin, 1989
\item{[ DT]} P. Deift, E. Trubowitz, Inverse scattering on the line, Comm. Pure Appl. Math. 32,
1979, 121-251.
\item{[  F]} L.D. Faddeev, Inverse problem of quantum scattering theory. II, Itogy Nauki i Tekh.
Ser. Sovrem. Probl. Mat. 3, 1974, 93-180 (in Russian). English translation in J. Sov. Math. 5, 1976, 334-396.
\item{[ GS]} F. Gesztesy, B. Simon, Inverse spectral analysis with partial information
on the potential. I. The case of an a.c. component in the spectrum, Helv. Phys. Acta 70, 1997, 66-71.
\item{[ HN]} G.M. Henkin, R.G. Novikov, Oscillating weakly localized solutions of the Korteweg-de Vries equation,
Teoret. Mat. Fiz. 61(2), 1984, 199-213 (in Russian); English translation: Theoret. and Math. Phys. 61(2), 1984,
1089-1099.
\item{[  K]} M.V. Klibanov, Phaseless inverse scattering problems in three dimensions,
SIAM J. Appl. Math. 74, 2014, 392-410.
\item{[KR1]} M.V. Klibanov, V.G. Romanov, Reconstruction formula for a 3-d phaseless inverse
scattering problem for the Schr\"odinger equation, arXiv:1412.8210v1, December 28, 2014.
\item{[KR2]} M.V. Klibanov, V.G. Romanov, Explicit formula for the solution of the phaseless
inverse scattering problem of imaging of nano structures, J. Inverse Ill-Posed Probl., DOI 10.1515/jiip-2015-0004
\item{[ KS]} M.V. Klibanov, P.E. Sacks, Phaseless inverse scattering and the phase problem in
optics, J.Math. Physics 33, 1992, 3813-3821.
\item{[  L]} B.M. Levitan, Inverse Sturm-Liuville  Problems, VSP, Zeist, 1987.
\item{[  M]} V.A. Marchenko, Sturm-Liuville Operators and Applications, Birkh\"auser, Basel, 1986.
\item{[ N1]} R.G. Novikov, Inverse scattering up to smooth functions for the Schr\"odinger equation
in dimension 1, Bull. Sci. Math. 120, 1996, 473-491.
\item{[ N2]} R.G. Novikov, Explicit formulas and global uniqueness for phaseless inverse
scattering in multidimensions, J. Geom. Anal. DOI:10.1007/s12220-014-9553-7;
\item{     }  arXiv:1412.5006v1, December 16, 2014.
\item{[ N3]} R.G. Novikov, Formulas for phase recovering from phaseless scattering data at fixed frequency,
Bull.Sci.Math. (to appear); arXiv:1502.02282v2, February 14, 2015.
\item{[ NM]} N.N. Novikova, V.M. Markushevich, On the uniqueness of the solution of the inverse
 scattering problem on the real axis for the potentials placed on the positive half-axis,
 Comput. Seismology 18, 1985, 176-184 (in Russian).

\end